\documentstyle[11pt,moriond,epsfig]{article}

\def\Journal#1#2#3#4{{#1} {\bf #2}, #3 (#4)}

\def\NPB{{\em Nucl. Phys.} B}
\def\PLB{{\em Phys. Lett.}  B}

\def\PRD{{\em Phys. Rev.} D}
\def\ZPC{{\em Z. Phys.} C}

\def\beq{\begin{equation}}
\def\eeq{\end{equation}}
\def\bea{\begin{eqnarray}}
\def\eea{\end{eqnarray}}

\def\to{\rightarrow}

\newcommand {\bi}      {\begin{itemize}}
\newcommand {\ei}      {\end{itemize}}
\newcommand {\ben}      {\begin{enumerate}}
\newcommand {\een}      {\end{enumerate}}
\newcommand {\bd}      {\begin{description}}
\newcommand {\ed}      {\end{description}}

\newcommand {\bc}      {\begin{center}}
\newcommand {\ec}      {\end{center}}

\newcommand {\emiss}  {\mathrm{E_{miss}}}
\newcommand {\jet}      {\mathrm{j}}

\newcommand {\chiuno} {\tilde{\chi}^0_1}

\newcommand {\grav}   {\tilde{\mathrm{g}}}
\newcommand {\sel}   {\tilde{\mathrm{e}}}

\newcommand {\slepton}   {\tilde{\mathrm{l}}}

\newcommand {\ee}{\mathrm{e}^+\mathrm{e}^-}

\newcommand {\G}{\mathrm{G}}

\begin{document}
\vspace*{4cm}
\title{EXOTIC SEARCHES AT LEP}

\author{ A. DE MIN }

\address{Dipartimento di Fisica, Universit\`a di Padova and INFN,\\
Via Marzolo 8,\\
Padova, Italia}

\maketitle\abstracts{
In the last five years, 
the operation of the LEP accelerator has
provided the HEP community with a unique opportunity to search for physics
beyond the standard model at the typical energy scale of $~200$ GeV.
Although most of the data analysis has been focused on the traditional
sectors of Higgs and gravity-mediated SUSY physics,
a considerable effort has 
been dedicated by ALEPH, DELPHI L3 and OPAL to the search for 
more exotic physics scenarios, which go from 
gauge-mediated supersymmetry breaking, to compositeness, 
to gravitational extra-dimensions. 
This talk reviews the latest experimental results
on a selection of final states which are believed to 
be of particular interest for exotic physics 
or which present some anomaly in the
data with respect to standard model expectations.}

\section{Introduction: searches for new physics at LEP}
At LEP the strategy adopted to look for signals of new physics
can vary, depending on the physics case,
between two extreme approaches.

One extreme possibility is to 
perform what is normally called an {\it indirect} search, which consists in
measuring with the best possible accuracy some reference standard  
model processes and comparing the obtained cross sections with the 
standard model expectations. 
Any disagreement beyond the experimental uncertainty
is interpreted as evidence for new physics. This approach
does not require the definition
of a precise theoretical framework alternative to the  
standard model, but 
it offers poor experimental sensitivity
(a few hundred fb for LEP2 luminosities)
because these processes
have typically large cross sections in the standard model.
Examples of this search strategy are the precise measurements of 
di-fermion or di-boson production cross sections, 
with or without initial state photons.


In order to improve the statistical sensitivity
a better  signal/background ratio is needed,
which in turn requires a more precise
definition of the signal to search for,
with a consequent loss of generality.  
This approach is followed, for example, 
in supersymmetry (SUSY) searches, 
where the signal sensitivity reaches the
level of a few tens of fb at LEP, 
but the result is strictly valid only in the context of the chosen
theoretical framework.
The very extreme case of this {\it direct} approach is the search for the
standard model Higgs, for which the theoretical model is 
completely defined, apart from the Higgs mass.
Here very specific analyses have reached sensitivities of some
10 fb.

Despite the fact that most searches have focused on
the SM Higgs and MSSM-SUGRA sectors, 
a considerable effort has been dedicated by ALEPH, DELPHI, L3 and OPAL to
the search for more exotic scenarios, which go from 
gauge-mediated supersymmetry breaking, to compositeness, 
to gravitational extra-dimensions. 
Since an exhaustive review of all these results would be impossible
in this context, only a selection of a few final states will
be discussed in detail, which are believed to be particularly
powerful in the search for new physics or that present
some anomaly in the data compared to standard model expectations.
For this purpose,
Table~\ref{tab:final_states} has been prepared which describes the 
correlation between a given final state and the  
exotic models that this final state can probe.
The table shows that, indeed, 
the simplest final states, such as $\ee \to f \overline{f} (\gamma)$
or $\ee \to \gamma (\gamma) + E_{miss}$,
can probe the largest number of models.
These and 
some other channels will be discussed in the following
sections.

{\footnotesize
\begin{table}[htb]
\renewcommand\arraystretch{1.0}
\caption{Sensitivity of the final states investigated at LEP 
for exotic physics.}
\begin{center}
\begin{tabular}{r|c|c|c|c|c|c|c|c|c|c|c|c|c}
 & \rotatebox{90}{ Contact Int.}                  
   & \rotatebox{90}{ Z$^\prime$}
     & \rotatebox{90}{ Extra-dim.} 
       &  \rotatebox{90}{  GMSB SUSY} 
        &  \rotatebox{90}{  Mass. sgoldstino}
          & \rotatebox{90}{  Spont. RpV SUSY}
           & \rotatebox{90}{ RpV SUSY} 
           & \rotatebox{90}{  Technicolor}       
             & \rotatebox{90}{  Excited fermions} 
               &  \rotatebox{90}{  Anom. coupl.}
                 &  \rotatebox{90}{  New fermions} 
                   &  \rotatebox{90}{  FCNC}
                    &  \rotatebox{90}{  LeptoQuarks} \\ \hline
  single-$\gamma$   &X&X& X& X& &X&& X& X& X  &&&\\
 Non-pointing single-$\gamma$   &&&&X&    &&&&&     &&&\\
  $\gamma \gamma \emiss$  &&&& X&    &&&& X& X &&&\\
 $\gamma \gamma (\gamma)$       &X&&X&&X   &&&&X&X    &&& \\
  ll($\gamma$)              & X& X& X&   &&& X& X& X &&&\\
 jj($\gamma$), gg$\gamma$       &X&X&X&&X &&&X&&X     &&&\\
 $\tau \tau \emiss$         &&&&X&    &X&&&&    &&& \\
 jjl $\emiss$                &&&&&     &&&&X&    &X&&X \\
  jjjj, bjjj, bbjj          &&&&&     &&X& X&&    && X& \\
 bjl $\emiss$                &&&&&     &&X&&&     &&X&X \\
 jj $\emiss$                  &&&&&     &&&&&     &&&X \\
 ll $\emiss$                  &&&&X&    &&&&X&    &&& \\
 jjll                          &&&&&    &&&&&     &&&X \\
 WW($\gamma$), ZZ, Z$\gamma$    &&&X&&     &&&X&&X    &&&\\
 Z $\emiss$                    &&&X&&     &&&&&    &&&\\
 multi-l, multi-j                &&&&&     &&X&&&    &&&\\
 ll$\gamma \gamma\ \emiss$    &&&&X&    &&&&X&    &X&&\\
 multi-l $\emiss$             &&&&X&    &&&&&     &&&\\
 ``kinked'' tracks              &&&&X&    &&&&&    &X&&\\
 heavy stable ptcs               &&&&X&    &&&&&    &X&&\\
 \end{tabular}
 \label{tab:final_states}
 \end{center}
 \end{table}
 }

Unless differently specified, the results mentioned in the this paper
refer to the data collected by ALEPH, DELPHI, L3 and OPAL
in the year 2000 at centre-of-mass energies ranging from 
202 GeV to 209 GeV and with a total integrated luminosity of
about 220 pb$^{-1}$ per experiment.
All  results are preliminary.

\begin{figure}
\begin{tabular}{cc}
\psfig{figure=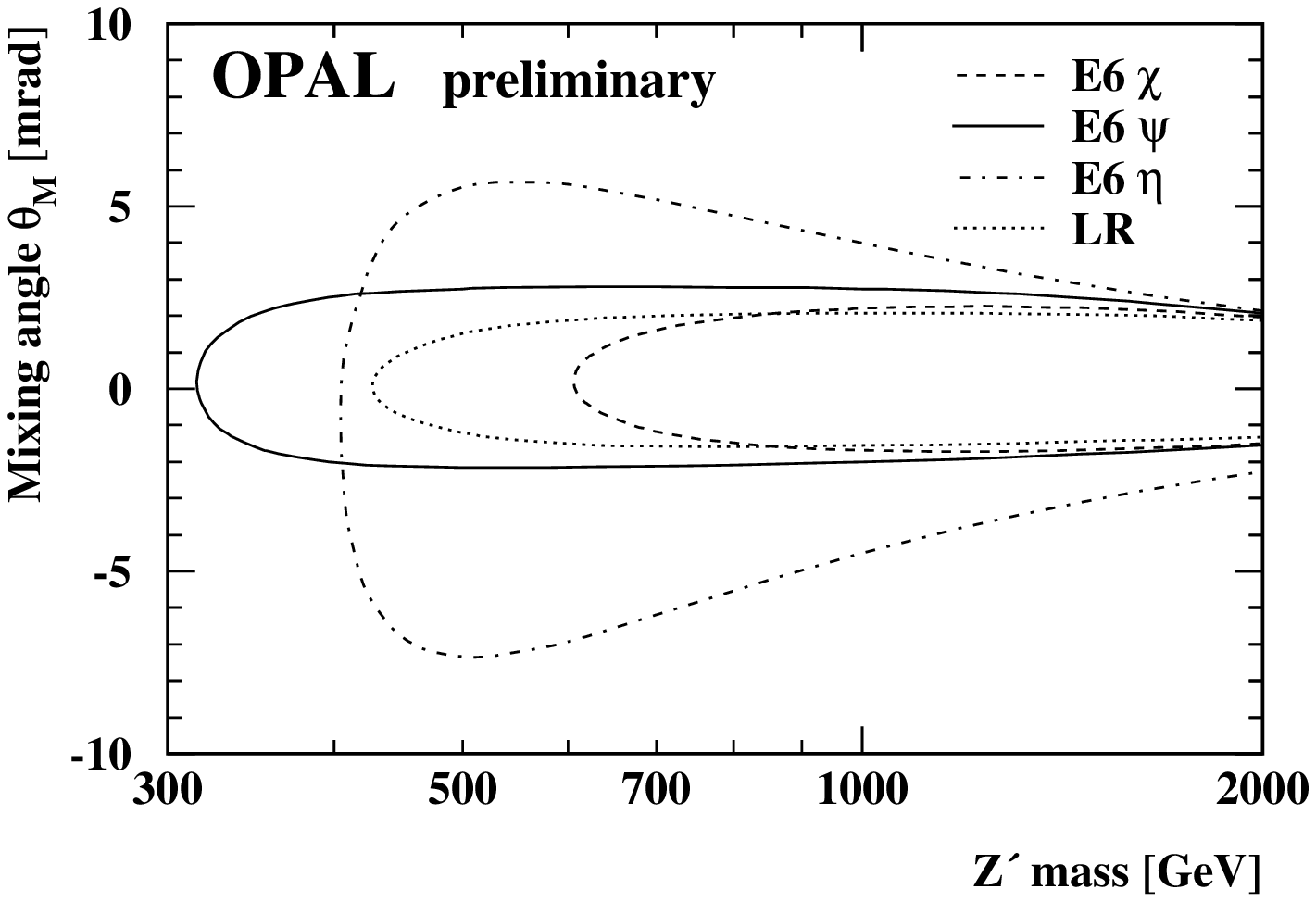,width=.5\textwidth} &
\psfig{figure=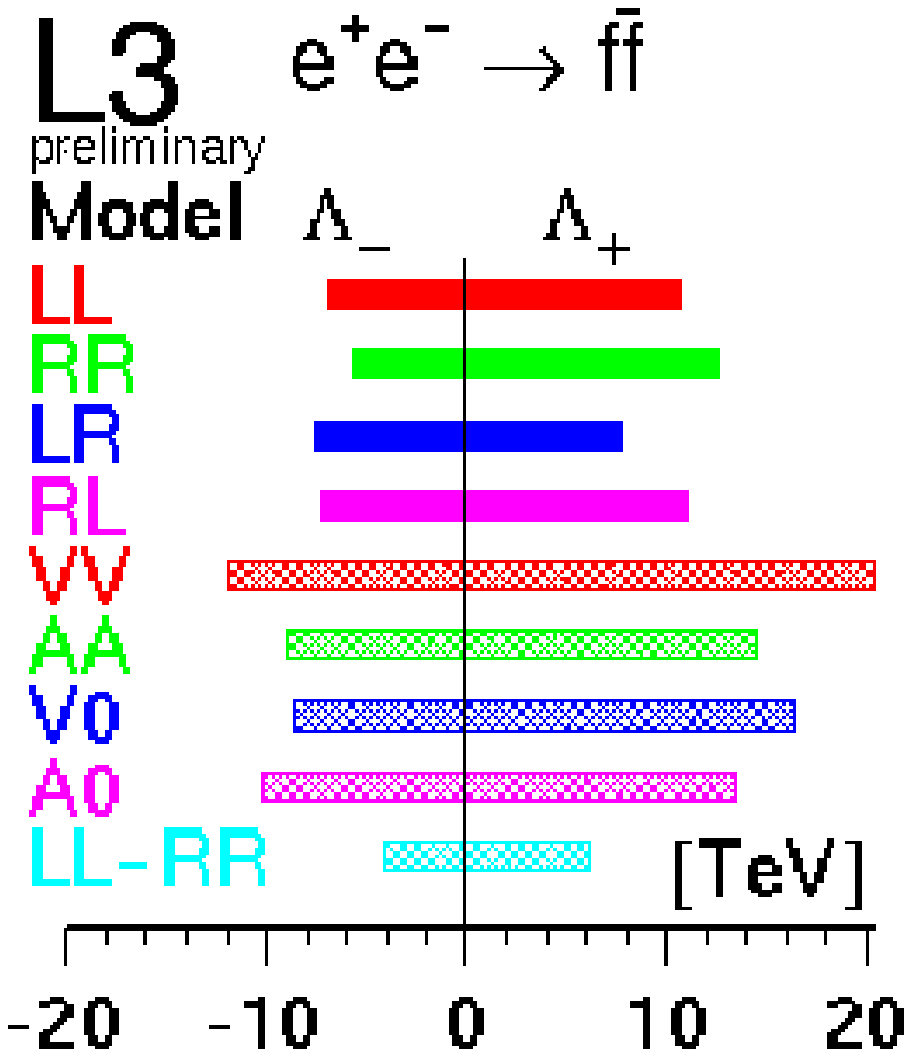,width=.3\textwidth}
\end{tabular}
\caption{Left: OPAL results on the search for Z$^\prime$. Right: 
L3 limits on 4-fermion contact terms in various 
theoretical frameworks.}
\label{fig:zprime}
\end{figure}

\section{Indirect searches with $\ee \to l^+l^- (\gamma)$ final states}
The most important reference process for indirect searches is the 
charged lepton pair-production, possibly accompanied by 
initial state radiation:
$\ee \to l^+l^- (\gamma)$.
Anomalies in the measured
cross section can indicate the existence of new interactions between
the initial and final state fermions, such as the exchange of
new heavy particles like
new vector bosons (Z$^\prime$), gravitons ($\G$) 
or heavy sneutrinos ($\tilde{\nu}$).
When the exchanged particles are {\it extremely} heavy,
the interaction becomes essentially point-like ({\it contact-interaction}), 
and is described by effective lagrangian terms of the type:
\begin{equation}
{\cal L} \approx {g^2 \over \Lambda^2} (\overline{e} \gamma ^\mu e )(\overline
f \gamma_\mu f).
\label{eq:contact}
\end{equation}
In equation~\ref{eq:contact} the $\Lambda$ term (squared) 
in the denominator is 
necessary to compensate the dimension-six operator of the 4-fermion 
interaction and describes, in some universal way, 
the typical energy scale of the new interaction, 
in exactly the same way as the Fermi constant $G_f^{-1/2}$ 
represents the energy 
scale of weak interactions in the Fermi model.

No evidence for anomalies has been found in the data collected by
LEP in the year 2000, as well as in the samples collected in the previous
years. Only OPAL~\cite{opalll} reports of a possible $\ee \to \tau^+ \tau^-
 \gamma$ excess at $\sqrt{s}>200$ GeV at a level of 2.5 $\sigma$.
The excess is not confirmed by the other 
experiments~\cite{alephll,delphill,l3ll} and should be
interpreted as a statistical fluctuation.
Limits have been derived on the $Z^\prime$ mass as a function of the
mixing angle with the standard model Z
and on the $\Lambda$ scale associated to 4-fermion contact-interactions. 
Some OPAL~\cite{opalll} and L3~\cite{l3ll} 
results are shown in Figure~\ref{fig:zprime};
the other experiments present similar results~\cite{alephll,delphill}.

\section{Search for new physics in single-photon events}
 
In the standard model single-photon events are produced mainly via
the process $\ee \to \nu \overline{\nu} \gamma$, where the photon is emitted
from the initial state electrons or 
from the W exchanged in the t-channel
(for  $\ee \to \nu_e \overline{\nu_e} \gamma$ only).
Compared to the study of charged fermion-pair production discussed in the
previous section, single-photon events present two distinctive features:
\begin{enumerate}
\item neutrinos cannot be produced via s-channel photon exchange
so the standard model
cross section is lower than $\ee \to f^+ f^- \gamma$ in some
phase space regions;
\item the final state is purely neutral and is thus sensitive to the 
production of new stable neutral particles X via $\ee \to X \gamma$ 
or $\ee \to XX \gamma$ processes (majorons, heavy neutrinos, gravitons, 
gravitinos, neutralinos,...).
\end{enumerate}

\begin{figure}
\begin{tabular}{cc}
\psfig{figure=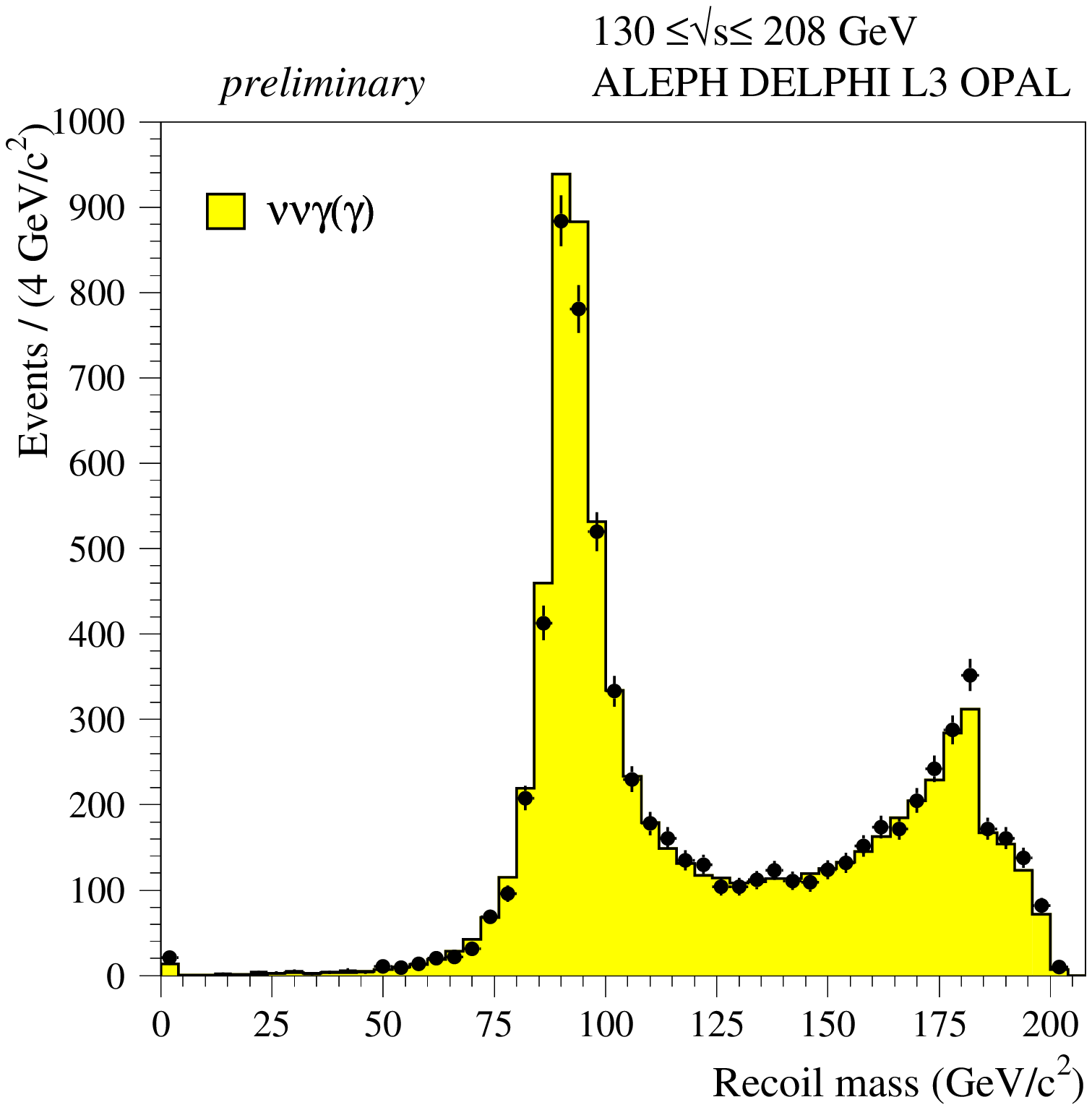,width=.45\textwidth}
\psfig{figure=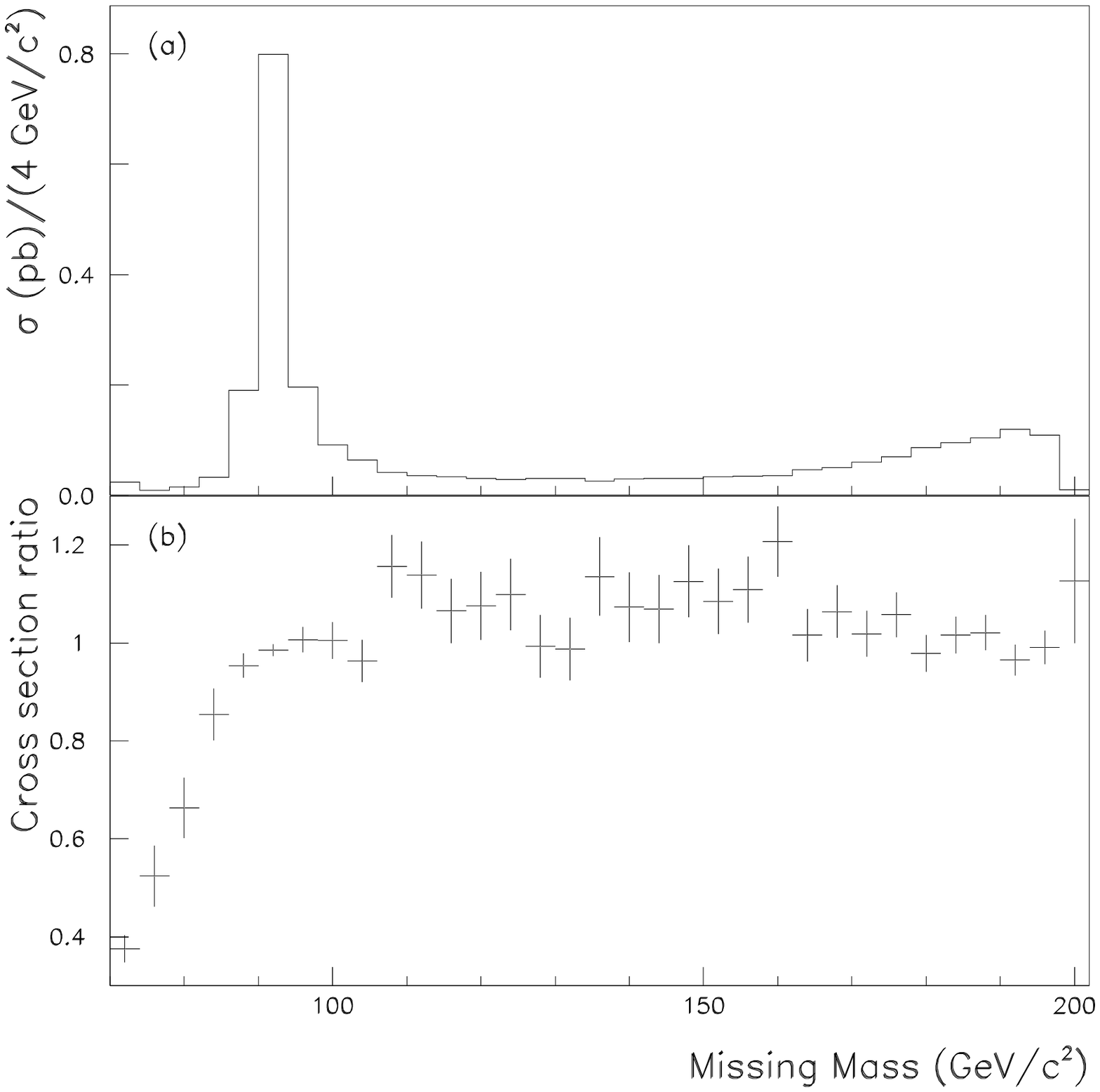,width=.45\textwidth}
\end{tabular}
\caption{Left: LEP2 combined 
missing mass distribution in single-photon events.
Right: Ratio between KK and Nunugpv expectations.}
\label{fig:single_photon}
\end{figure}

Deviations can be searched for in the photon energy and angular
distributions. For example, some SUSY models with 
light gravitinos~\cite{gravheavy}
and some models with gravitational extra-dimensions~\cite{extradim} predict 
an excess of
events with low photon energy (particularly at small polar angle).


The combined single-photon missing-mass spectrum observed by the four LEP collaborations
in all LEP2 data ($130 \leq\sqrt{s}\leq 209$ GeV)~\cite{susywg} is shown in 
Figure~\ref{fig:single_photon}, where it is compared to a simulation of the standard model
background. Discrepancies are visible 
below the Z peak, where a deficit of some 5\% is seen,
and at very low photon energies, where an excess of some 5\% is observed.
Unfortunately, these results cannot be interpreted 
as evidence of new physics such as
GMSB or extra-dimensions, since the {\it shape}
of the standard model simulation shown in Figure~\ref{fig:single_photon}
is affected by a systematic uncertainty of order 5\%,
evaluated on the basis of the discrepancies
between the two most used 
generators: Koralz~\cite{koralz} and Nunugpv~\cite{nunugpv}.
More detailed investigations show that:
\begin{enumerate}
\item the deviation is observed consistently 
by all four experiments, although it
is particularly evident in L3 results~\cite{l3_photons};
\item the deviation does not have an evident dependence on the 
centre-of-mass energy;
\item the Nunugpv generator seems to fit the data fairly well, while the
largest disagreement is observed with respect to Koralz.
\end{enumerate}
A third generator, KK~\cite{kk}, not publicly available a the time 
of the conference,
will soon replace Koralz. A preliminary comparison with
Nunugpv, performed by the ALEPH Collaboration~\cite{aleph_photons}, shows
a better agreement than between Koralz and Nunugpv, apart from
a limited region immediately below the Z return peak 
(Figure~\ref{fig:single_photon} Right). 

For this channel, final conclusions will
be drawn only after KK-based simulations will be available
to all four collaborations. 

\section{Multi-photon final states}

\begin{figure}
\begin{tabular}{cc}
 \psfig{figure=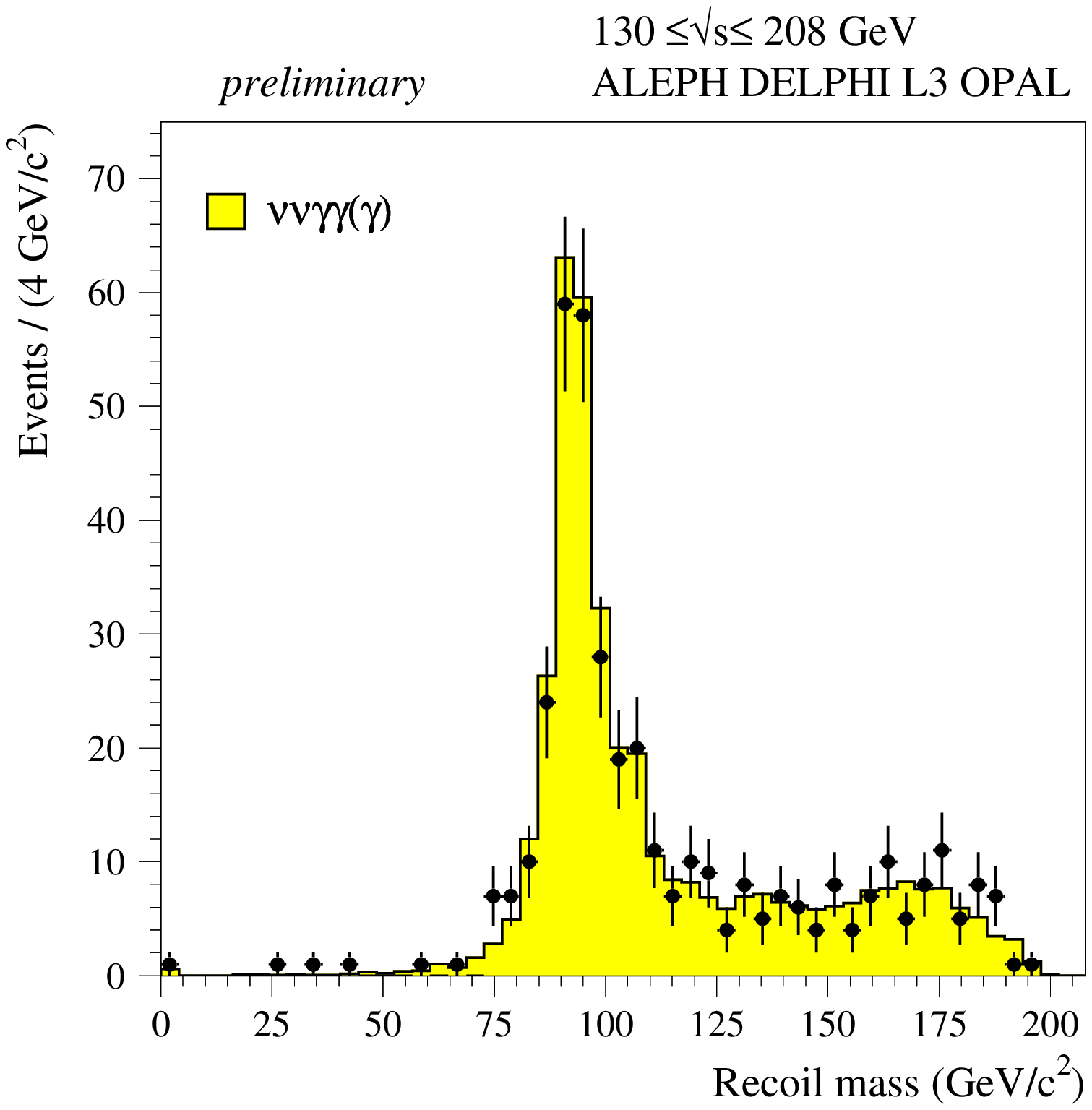,width=.45\textwidth}&
 \psfig{figure=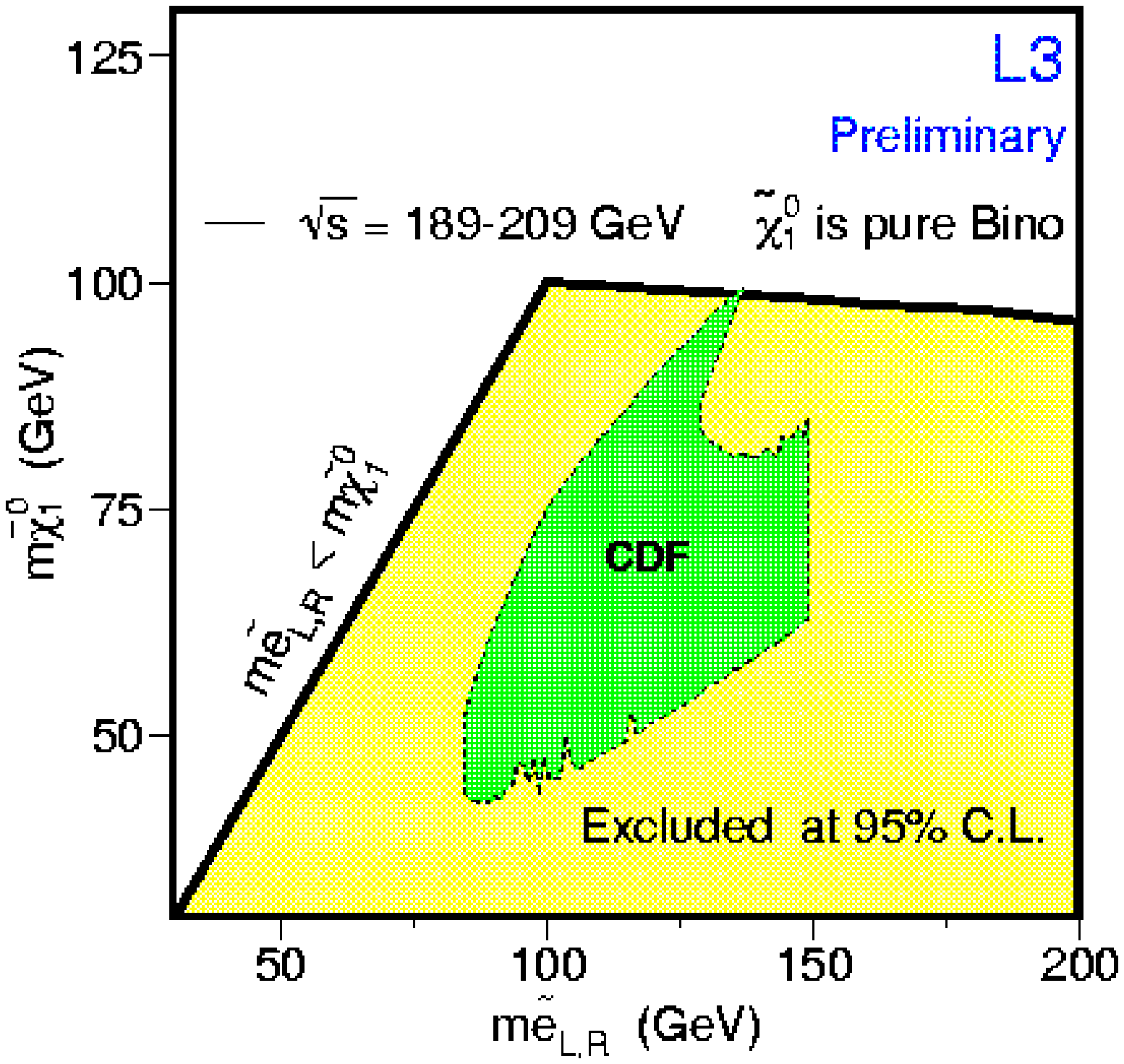,width=.45\textwidth}
\end{tabular}
\caption{
Left:Combined LEP missing mass distribution in events with acoplanar photon pairs.
Right:
Exclusion limit obtained by the L3 Collaboration in the 
(m$_{\chiuno}$, m$_{\sel}$) plane assuming 
BR($\chiuno \to \grav \gamma$)=100\%. The region allowed by the
CDF ee$\gamma \gamma \emiss$ event is now excluded.
}
\label{fig:acoplanar}
\end{figure}

One particular multi-photon final state has revealed to be of
particular importance for exotic searches at LEP: 
$\ee \to \gamma \gamma \emiss$, 
often address as ``acoplanar-photons''.
This is the ideal channel to look 
for evidence of GMSB~\cite{gmsb}
with neutralino-NLSP via the process
$\ee \to \chiuno \chiuno \to \grav \gamma \grav \gamma$.
The light gravitinos 
escape detection and the photons appear as acoplanar.
Other exotic processes investigated with the same final 
states are, for example,
the presence of anomalous 4-vector boson couplings
and excited neutrino pair-production (in case of $\nu^* \to \nu \gamma$ decay).

In this channel no anomalies are observed with respect to 
Koralz or Nunugpv (Figure~\ref{fig:acoplanar} Left), 
given the large statistical uncertainty affecting the data, 
and limits on new physics are derived.
The limit obtained by the L3 Collaboration on the $\chiuno$ 
mass~\cite{l3_photons} is shown in 
Figure~\ref{fig:acoplanar} (Right) as a function of the 
selectron mass (which determines the 
$\ee \to \chiuno \chiuno \to \grav \gamma \grav \gamma$ cross section
via t-channel exchange) and excludes, once for all,
the GMSB scenario suggested 
by one $ee\gamma \gamma \emiss$ event collected
by the CDF experiment at Fermilab in 1996~\cite{cdfevent}.
Similar limits are obtained by the other LEP collaborations~\cite{aleph_photons,delphi_photons,opal_searches}.

\begin{figure}[t]
\begin{tabular}{cc}
\psfig{figure=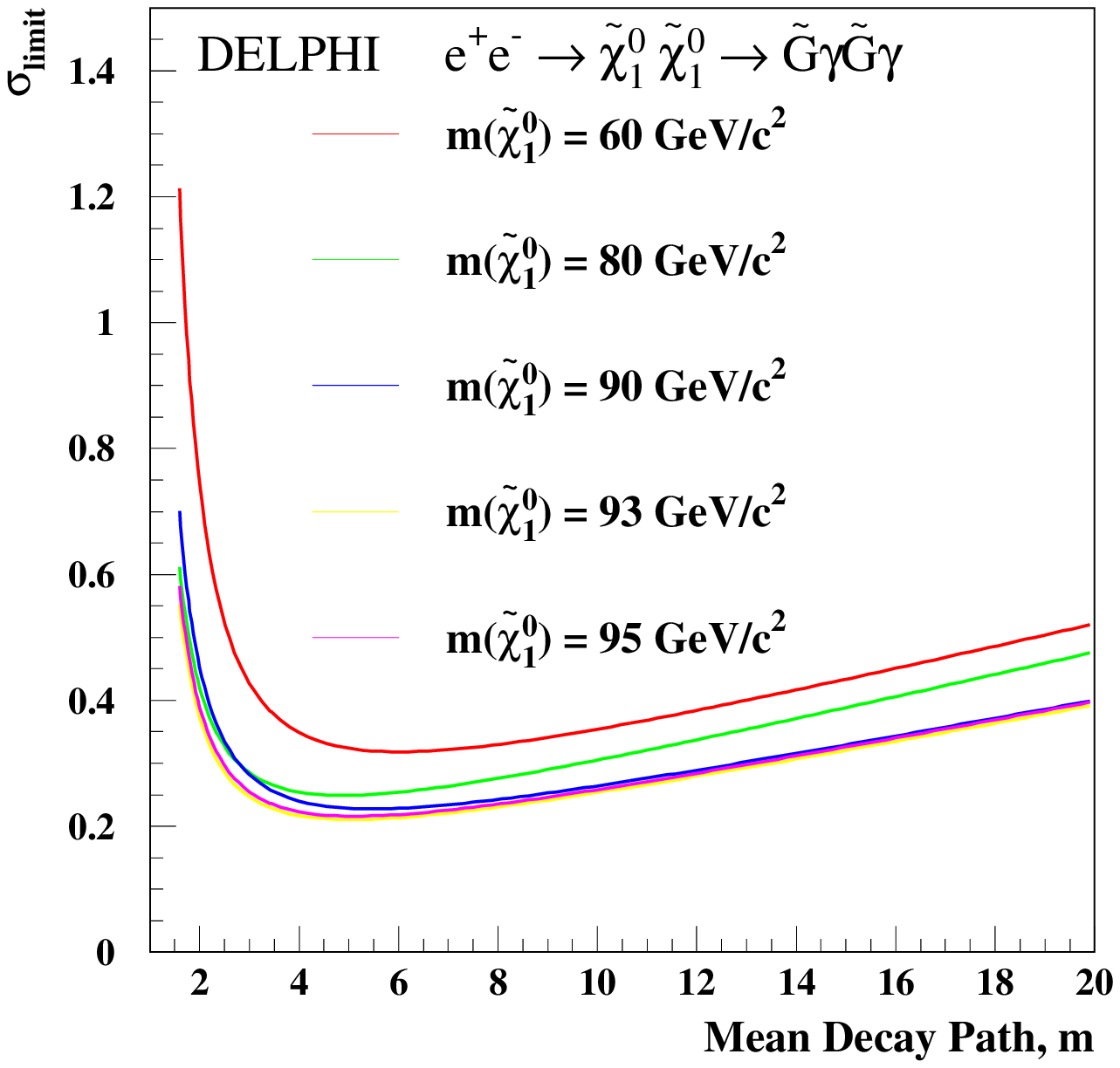,width=0.42\textwidth}&
\psfig{figure=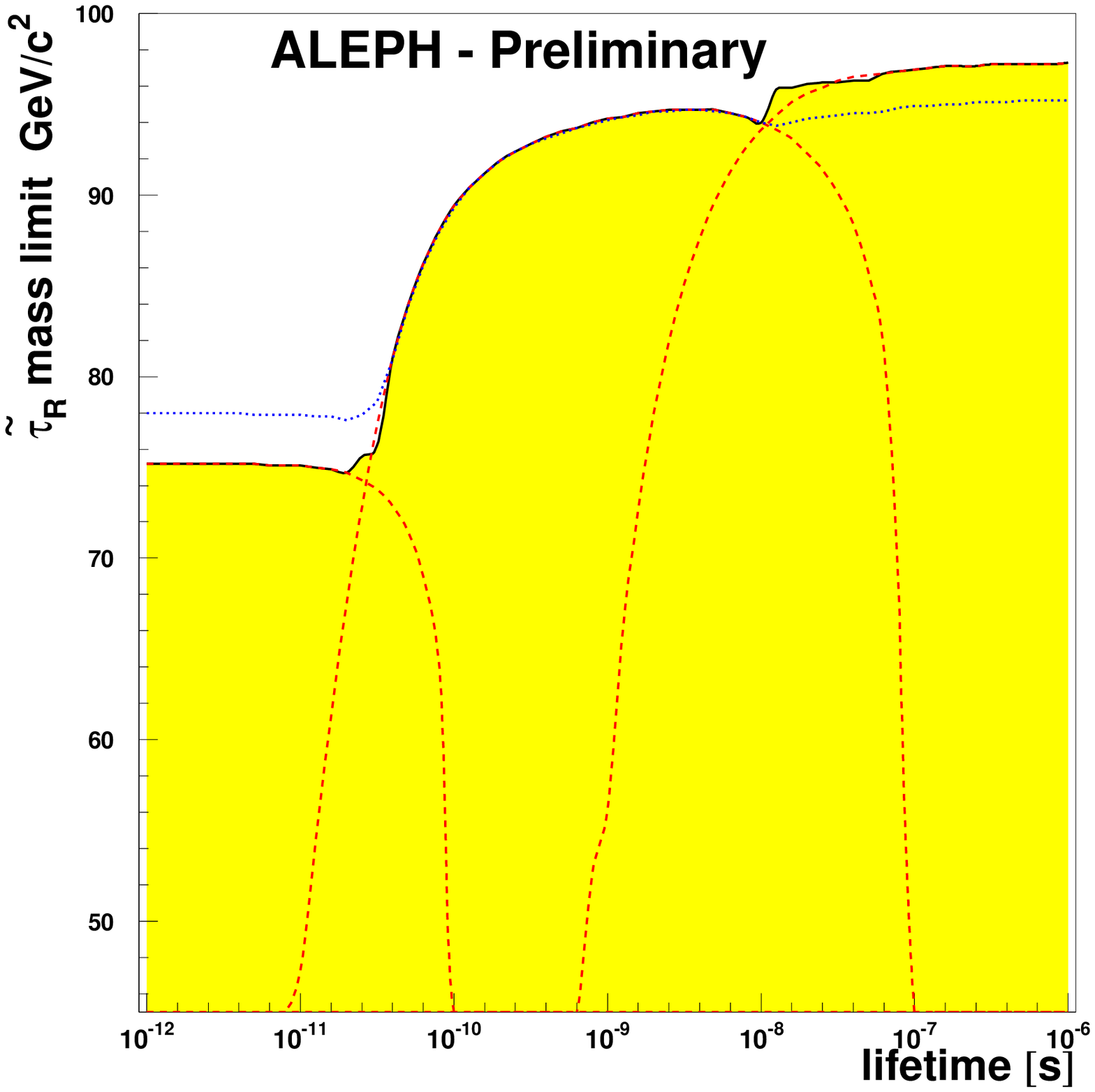,width=0.43\textwidth}
\end{tabular}
\caption{Left: Exclusion limit obtained by the DELPHI Collaboration
on $\sigma(\ee \to \chiuno \chiuno)$ as a function of the $\chiuno$ decay
path.
Right: Aleph exclusion limit on the stau-NLSP mass in GMSB as a 
function of the stau lifetime.
}
\label{fig:kinks}
\end{figure}

The result in Figure~\ref{fig:acoplanar} (Right) is valid for light gravitinos,
with masses of order 100 eV or smaller. For larger $\grav$ masses, the
decay $\chiuno \to \grav \gamma$ would have measurable lifetime and would
produce events with photons which do not
originate from the beam interaction region. As an example, the
DELPHI cross-section limit for this process is shown as a function of the 
neutralino mean decay path in Figure~\ref{fig:kinks} (Left).

\section{Acoplanar (kinked) leptons}

Searches for GMSB in case of lepton-NLSP are based on the analysis of
events with acoplanar lepton pairs. As for the $\chiuno$-NLSP case the
lifetime $\tau(\slepton \to l \grav )$ is a function of 
the gravitino mass squared. Depending on $m_{\grav}$ one
expects final states with simple acoplanar leptons ($\tau < 1$ mm),
tracks with visible impact parameter (1 mm$<\tau < 1$ cm),
tracks with a decay kink in the tracking devices (1 cm$<\tau < 1$ m), or
heavy stable charged particles ($\tau >$1 m).


No evidence for any of these processes has been observed at LEP.
The limit obtained by ALEPH~\cite{aleph_photons} on the stau mass 
as a function of the
stau lifetime is shown in Figure~\ref{fig:kinks} (Right).

\section{Final states with two jets and a photon}

Theories with light gravitinos also predict the existence of sgoldstinos
($\Phi$), the goldstino superpartners, which have arbitrary mass~\cite{sgold}.
Being superpartners of a supersymmetric particle, sgoldstinos have
even R-parity and do not need to be produced in pairs so the 
largest cross section is expected for the process $\ee \to \Phi \gamma$.
Sgoldstinos decay into gluons (dominant) or photons.

\begin{figure}[t]
\begin{tabular}{cc}
\psfig{figure=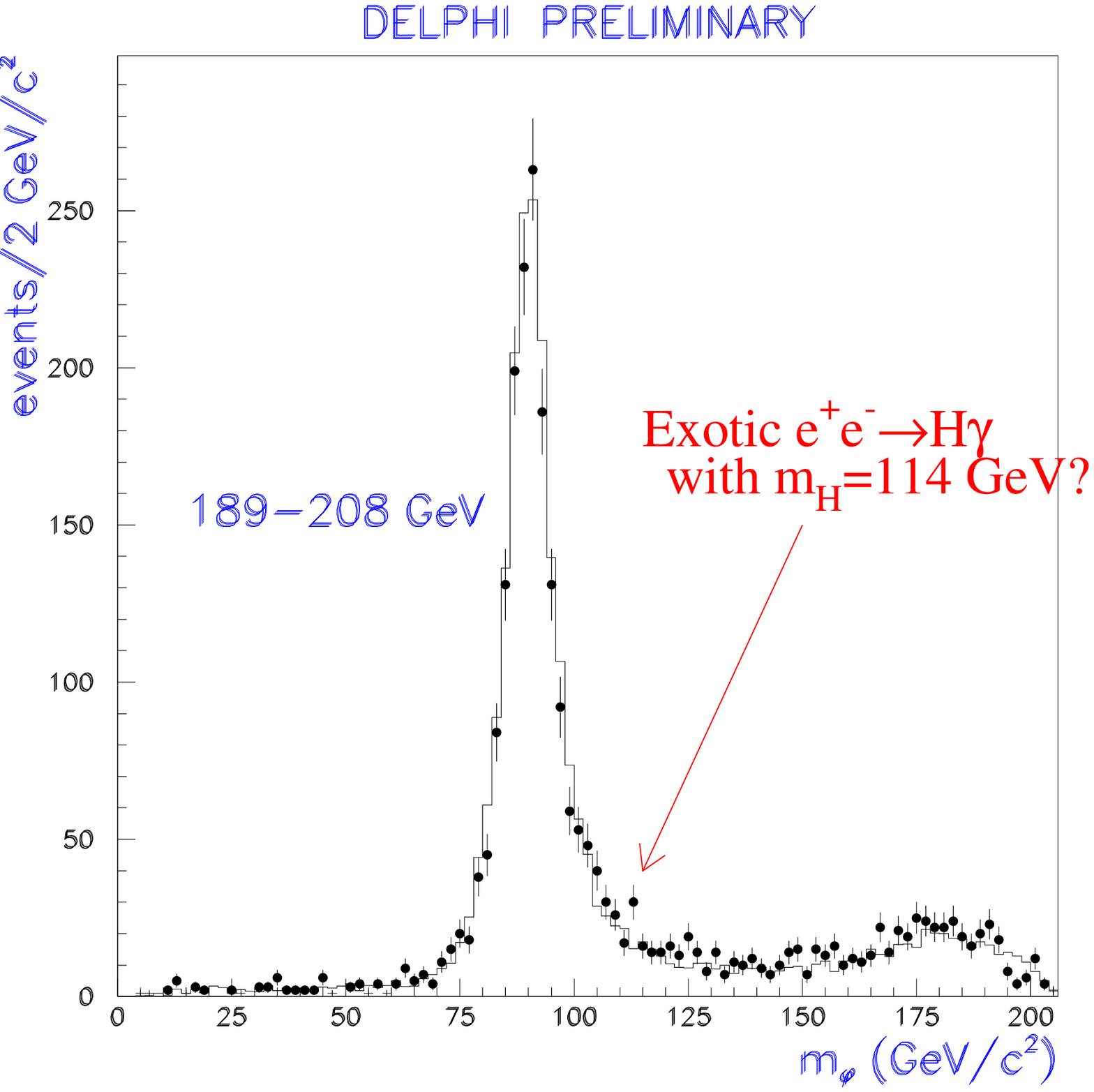,width=0.4\textwidth}&
\psfig{figure=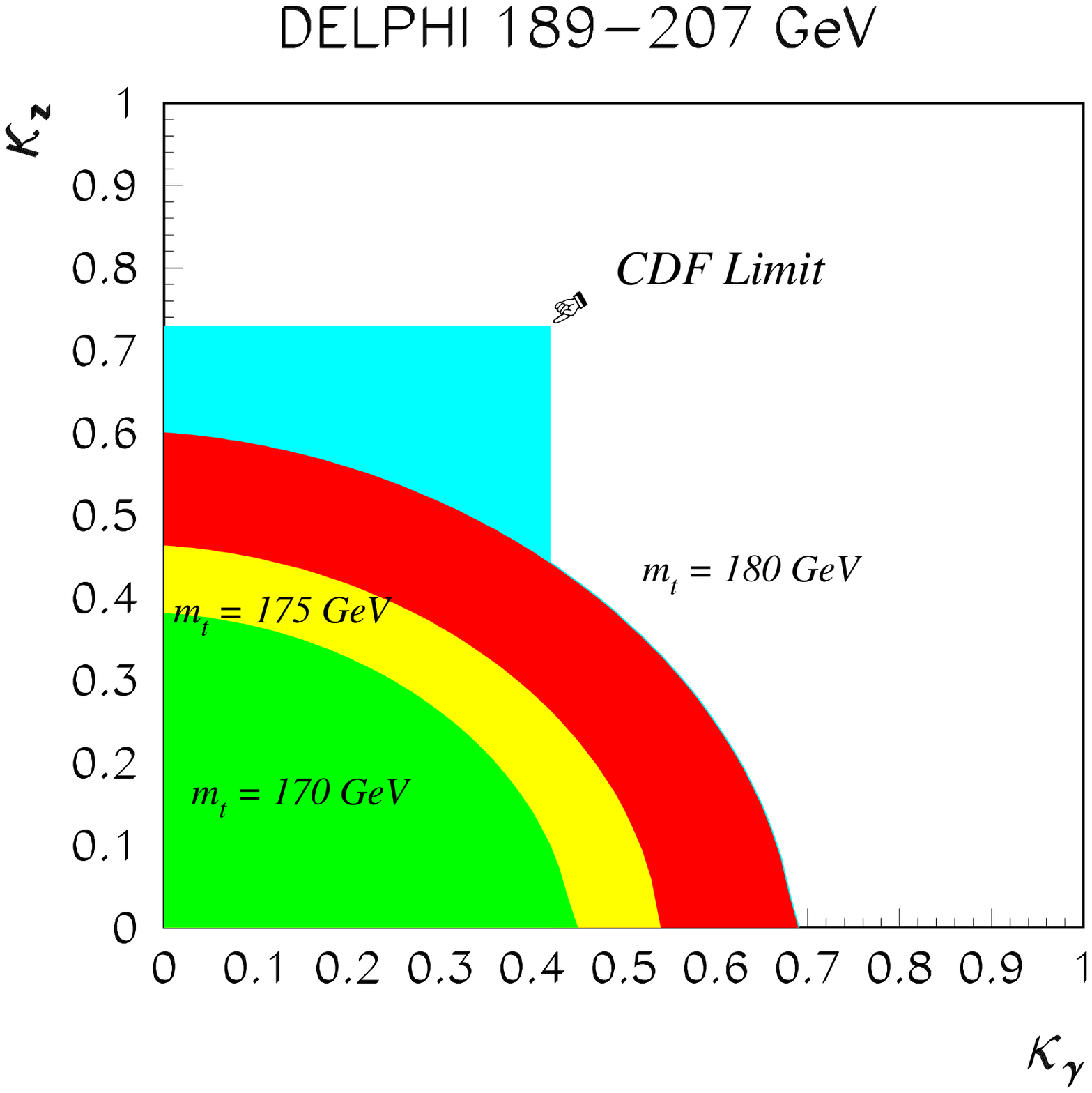,,width=0.48\textwidth}
\end{tabular}
\caption{Left: Invariant mass distribution of the two jets recoiling 
the isolated photon in jj$\gamma$ events in DELPHI.
Right: Exclusion limit in the ($k_\gamma$, $k_Z$) plane set by
DELPHI from the study of jbW final states (compared to a CDF limit).}
\label{fig:sgoldstino}
\end{figure}

A search for sgoldstino has been recently performed by the DELPHI 
Collaboration~\cite{delphi_sgold}. The key distribution is the mass recoiling the photon in
jet-jet-$\gamma$ final states ($\ee \to \Phi \gamma \to gg\gamma$), 
which is shown in Figure~\ref{fig:sgoldstino} (Left)
for the data sample $189<\sqrt{s}<208$ GeV. No evidence for a signal is 
observed. The tiny peak at m$_{jj}\sim 115$, 
which might indicate 
exotic Higgs production via $\ee \to H \gamma$,
was seen in 1999 data and not confirmed by 2000 data.

The same jj$\gamma$ final state can be used to search for 
Technicolor ($\ee \to \rho_T \gamma \to \jet \jet  \gamma$ or 
$\ee \to \phi^0_T \gamma \to \jet \jet \gamma$),
whose minimal version is excluded by LEP1~\cite{tech1} but
which still survives in non-minimal versions 
(walking technicolor)~\cite{tech2}.
Even in this second case, $\rho_T$ masses below 200 GeV are excluded by 
LEP2 data~\cite{delphi_4j}. 

\section{Four-jet final states}

\begin{figure}[t]
\begin{tabular}{cc}
\psfig{figure=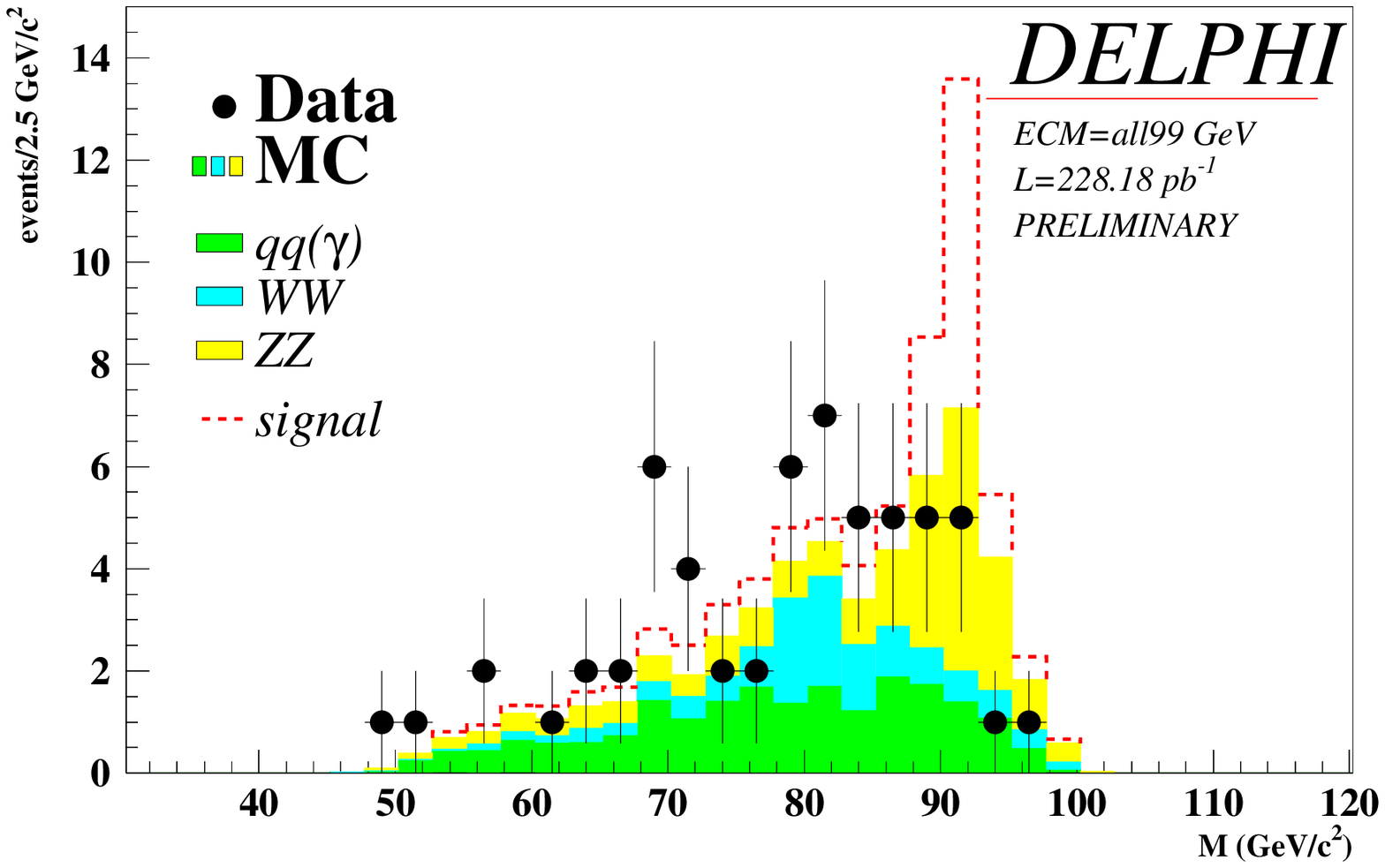,width=0.45\textwidth}&
\psfig{figure=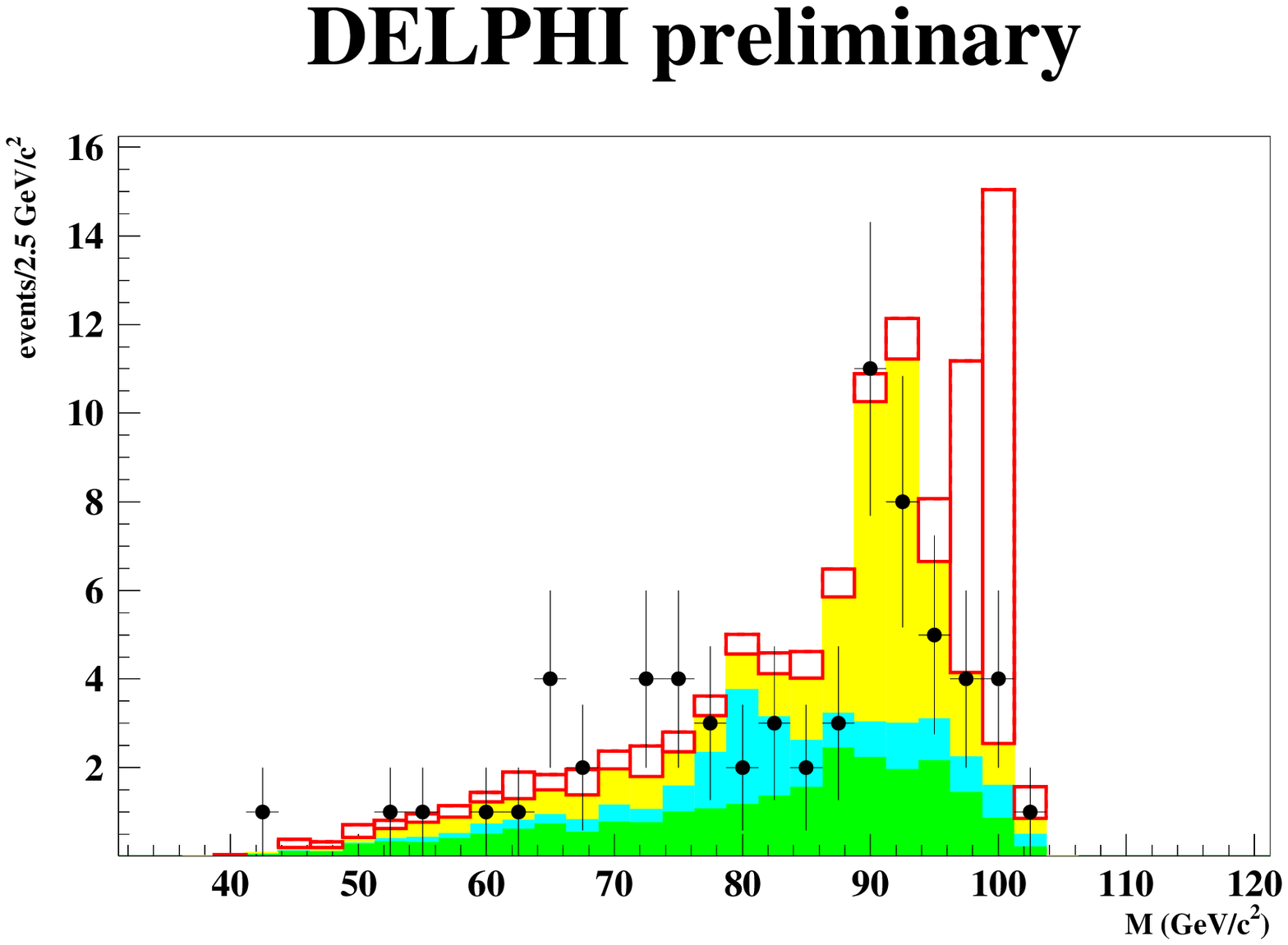,width=0.45\textwidth}\\
\psfig{figure=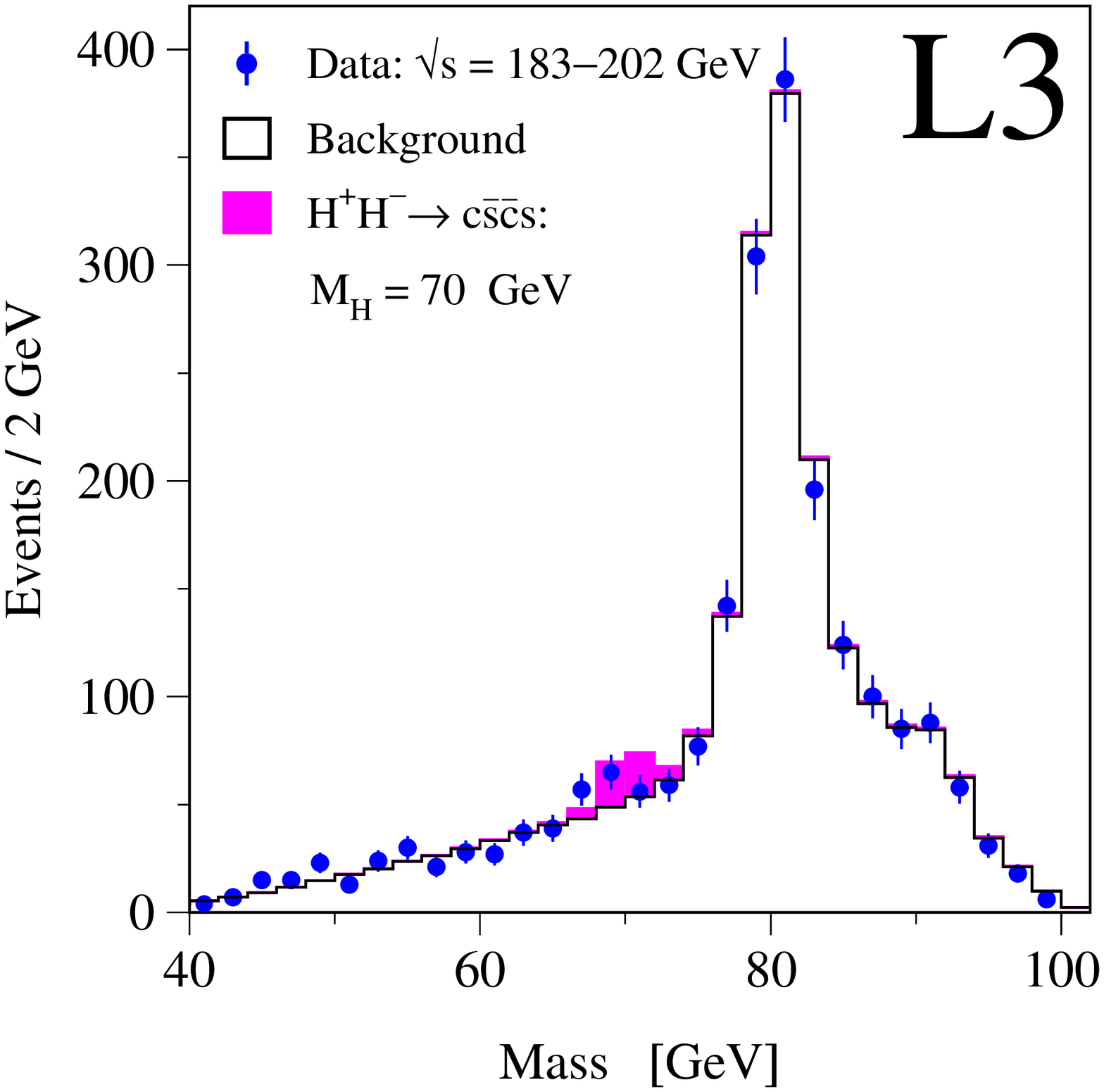,width=0.4\textwidth} &
\psfig{figure=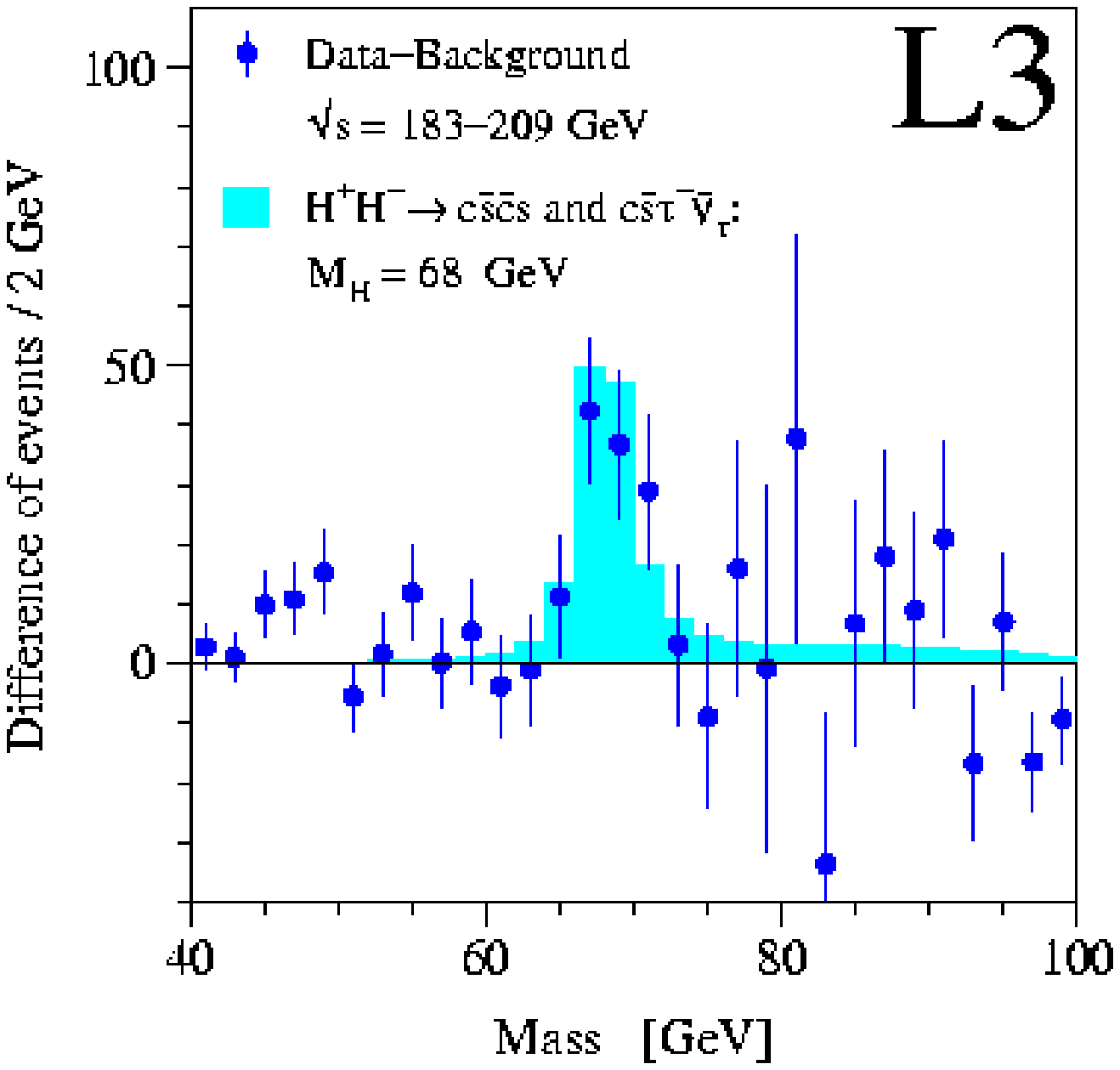,width=0.43\textwidth}
\end{tabular}
\caption{Top: The anomalous peak observed by DELPHI in 4-jet events in 1999 (Left) disappears with the 2000 update (Right).
Bottom: L3 confirms the 1999 data anomaly (Left) with 2000 data (Right).}
\label{fig:technicolor}
\end{figure}

To extend the LEP sensitivity for $\rho_T$ masses above 200 GeV 
the analysis of four-jet final states is used,
assuming $m_{\pi_T}<\sqrt{s}/2$.
Four-jet events are
probably the most deeply studied final state
at LEP2, given its importance for Higgs searches and
W physics. Still, LEP results in this topology have always 
been somehow controversial, especially for the difficulties associated
to the choice of the correct jet-pairing in the estimation of the jet-jet 
invariant masses.
The first anomalies in four-jet events were observed by ALEPH  
in 1995~\cite{aleph_4j_old},
but disappeared with larger statistics. 
In 1999 DELPHI~\cite{delphi_4j_old} and L3~\cite{l3_4j_old} also 
observed some (weak) 
excess over the background estimations in 4-jet events
having jet-jet invariant mass close to 68 GeV.
Although the analyses assume different physics
scenarios (Technicolor searches for DELPHI, charged Higgs searches for L3)
the final state is similar and the excess is localised
in the same jet-jet invariant mass region (Figure~\ref{fig:technicolor}). 
By adding the data collected
in 2000 the DELPHI anomaly disappears~\cite{delphi_4j}, 
while the L3 peak is confirmed and
reaches a final significance of 4.4 standard 
deviations~\cite{l3_4j}. 
Since ALEPH~\cite{aleph_4j} and OPAL\cite{opal_4j} do not
observe any anomaly with similar analyses, this L3 result has to 
be interpreted either as a statistical fluctuation, which is unlikely given
the significance, or as a spurious effect of the analysis.

\section{Final states with jets and leptons}

Among final states containing jets and leptons, 
the $\ee \to$jbW channel
deserves some comments since it is relatively new at LEP.
An excess in $\ee \to$jbW events could
indicate anomalous single-top production, via 
$\ee \to t\overline{u}/ t \overline{c}$ processes, as predicted
by several exotic scenarios, from models with anomalous Z couplings~\cite{single_top_coup},
to theories with dynamical ew-symmetry breaking~\cite{single_top_dyn}, 
to SUSY with R-parity violation (RpV).

Unfortunately, no evidence for any excess is observed in the data.
The results are thus expressed as limits for anomalous Z
couplings~\cite{delphi_top} 
(Figure~\ref{fig:sgoldstino} Right) or for the top
decay width via neutral currents: BR$(t \to Zu / Zc)<17\%$~\cite{aleph_top}. 
It's interesting to notice that, by exploiting the
kinematic features of the signal at LEP (the top is produced almost
at rest in the laboratory frame), the experimental sensitivity 
is comparable to that of Tevatron experiments.

\section{Multi-lepton, multi-jet final states}

Multi-particle final states are the ideal environment to search
for SUSY with R-parity non-conservation. 
The number of different event
topologies studied by the four LEP collaborations is impressive and
gives an idea of the global effort dedicated to the search for 
new physics. Indeed, excellent agreement is observed between
data and background expectations, with, maybe, one exception:
OPAL data in the 
$\ee \to \tau^+\tau^-+ \leq $(4 jets) channel at $\sqrt{s}\geq 206.5$ GeV~\cite{opal_searches}. 
Here 8 events are seen while 2.2 are expected from standard model
sources, which corresponds to a Poisson probability of $2 \cdot 10^{-3}$.
A similar analysis ($\ee \to \tau^+\tau^-+$4 jets)~\cite{aleph_rpv}, 
performed by the 
ALEPH Collaboration, yields 6 (9) events with a background of 8.0 (5.0) 
at $\sqrt{s}>205.5$ GeV ($\sqrt{s}<205.5$ GeV): if there's an excess 
this is concentrated at lower centre-of-mass energies according to ALEPH, 
which somehow contradicts the OPAL anomaly.

\section{Conclusions}
The last year of LEP operation has been extremely successful for what concerns
luminosity and beam energy:
more than 200 pb$^{-1}$ of data at centre-of-mass energies between
202 and 209 GeV have been collected, exceeding even the more optimistic 
expectations.
Despite the large statistics accumulated 
and the huge effort dedicated by ALEPH, DELPHI, L3 and OPAL 
to search for any possible discrepancy with the standard model 
in any final state topology, 
no real evidence for exotic physics has been found.
Only two channel still offer some matter of discussion, namely
single-photon events, where a better control of the theoretical
uncertainties is needed, and 4-jet final states, where
a possible signal of charged Higgs production reported by L3 is
not confirmed by the other experiments.

\end{document}